\documentclass[twocolumn,showpacs,showkeys,preprintnumbers,superscriptaddress,prl]{revtex4-1}
\usepackage{amssymb}

\usepackage{amsfonts}

\usepackage{latexsym}

\usepackage{dcolumn}

\usepackage{amsmath}

\usepackage{graphicx}

\newcommand{\eqn}{\ref}

\begin{document}

\title{Reversible optical to microwave quantum interface}

\author{Sh. Barzanjeh}
\affiliation{School of Science and Technology, Physics Division, University of Camerino, Camerino (MC), Italy}
\author{M. Abdi}
\affiliation{School of Science and Technology, Physics Division, University of Camerino, Camerino (MC), Italy}
\affiliation{Department of Physics, Sharif University of Technology, Tehran, Iran}
\author{G. J. Milburn}
\affiliation{Centre for Engineered Quantum Systems, School of Physical Sciences, The
University of Queensland, Saint Lucia, QLD 4072, Australia}
\author{P. Tombesi}
\affiliation{School of Science and Technology, Physics Division, University of Camerino, Camerino (MC), Italy}
\author{D. Vitali}
\affiliation{School of Science and Technology, Physics Division, University of Camerino, Camerino (MC), Italy}

\date{\today}

\begin{abstract}
We describe a reversible quantum interface between an optical and a microwave field using a hybrid device based on their common interaction with a micro-mechanical resonator in a superconducting circuit. We show that, by employing state-of-the-art opto-electro-mechanical devices, one can realise an effective source of (bright) two-mode squeezing with an optical idler (signal) and a microwave signal, which can be used for high-fidelity transfer of quantum states between optical and microwave fields by means of continuous variable teleportation.
\end{abstract}


\maketitle

Quantum technologies will achieve maturity only when it becomes possible to integrate distinct modules in a single "hybrid" device, achieving a functionality that transcends the capability of any one component~\cite{Wallquist2009}. In general, this will require a quantum interface, able to transfer coherently and faithfully quantum information between the modules, without introducing decoherence. A very useful interface would enable communication between superconducting
microwave systems and atomic-molecular-optical systems, or indeed between superconducting
systems in distinct low temperature environments \cite{Folman2001,Kubo2010,Schuster2010}.

A number of schemes for a quantum interface between light at different wavelengths have
been demonstrated \cite{Marcikic2003,Tanzilli2005,Rakher2010}, and very recently various solutions for interfacing
optics and microwaves have been proposed \cite{Matsko2007,Tian2010,Tsang2010,Regal2011,Tsang2011,Taylor2011,Wang2012,Tian2012,Winger2011,Hafezi2012}.
We describe here a reversible quantum interface
between optical and microwave photons based on a micro-mechanical resonator in a superconducting circuit, simultaneously interacting with an optical and a microwave cavity.

When the cavities are appropriately driven, the mechanical resonator mediates an effective parametric amplifier interaction, entangling an optical signal and a microwave idler. Such continuous variable (CV) entanglement can be then exploited to implement CV teleportation \cite{Braunstein1998}. The optical output is mixed with an optical 'client' field in an unknown quantum state on a beam splitter at the transmitting site (Alice).   The two outputs are then subject to  homodyne detection (see Fig.~1) and the classical measurement results communicated to the receiving site (Bob). Upon receipt of these results, Bob makes a conditional displacement of the microwave field, again using beam splitters and a coherent microwave source. The resulting state of the output microwave field is then prepared in the same quantum state as the optical input state. The process is entirely symmetric: the Alice and Bob roles can be exchanged and an unknown input microwave field can be teleported onto the optical output field at Alice, realising therefore a reversible quantum state transfer between fields at completely different wavelengths.


\begin{figure}[ht]
   \centering
  \includegraphics[width=.45\textwidth]{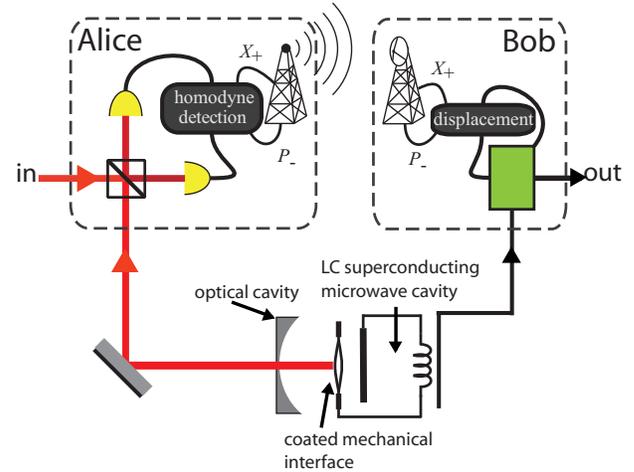} 
   \caption{Schematic description of the proposed optical-microwave interface. Alice mixes the optical cavity output with the input optical field she has to transfer, and communicates the results of her Bell measurements to Bob. The output state of the microwave field in Bob's hands is a faithful copy of the optical input field state when the optical-microwave cavity outputs are strongly entangled by their common interaction with the mechanical resonator, and Bob correctly displaces its state conditioned to Alice's measurement results. The scheme can be reversed by exchanging the roles of Alice and Bob: Bob performs the Bell measurements on microwave fields and the state of an input microwave field is teleported onto the optical output in Alice hands after communication of the measurement results and the conditional displacement.
   }
   \label{fig:1}
\end{figure}

 We assume a mechanical resonator (MR) which on the one side is capacitively coupled to the field of a superconducting microwave cavity (MC) of resonant frequency $\omega_w$ and, on the other side, coupled to a driven optical cavity (OC) with resonant frequency $\omega_c$ (see Fig.~1) . Such a device could be realized for example by adding an optical cavity to the superconducting circuit system of Teufel et al. \cite{Teufel2011}, by depositing an highly reflective coating on the drum-head capacitor of the circuit and driving it with a laser through a standard input mirror. Alternatively one could adopt a membrane-in-the-middle setup \cite{Thompson2008} in which a metal-coated membrane \cite{Yu2012} is capacitively coupled to a microwave cavity. The microwave and optical cavities are driven at the
frequencies $\omega_{0w}=\omega_w-\Delta_{w}$ and
$\omega_{0c}=\omega_c-\Delta_{c}$, respectively, where $\Delta_{j}$, $j=c,w$ are the respective detunings.

The single photon coupling constants between the optical and microwave resonator fields and the mechanical resonator are small in current experiments so we linearise the equations of motion by expanding around the
steady state field amplitudes in each resonator, $\alpha_s$ and $\beta_s$ (see Supplementary Information).  When $|\alpha_s|\gg 1$ and $|\beta_s|\gg 1$, and stability conditions are satisfied, the dynamics around the fixed point can be safely linearized and the effective Hamiltonian, in the interaction picture,  is given by~\cite{Vitali2007Milb,Barzanjeh2011a}
\begin{eqnarray}
H & = & \hbar\Delta_c\hat{a}^{\dagger}\hat{a}+\hbar\Delta_w \hat{b}^{\dagger}\hat{b}+\hbar\omega_m \hat{c}^\dagger \hat{c}\\\nonumber
& & -\hbar G_{c}(\hat{a}^\dagger+\hat{a})(\hat{c}+\hat{c}^\dagger) -\hbar G_{w}(\hat{b}^{\dagger}+\hat{b})(\hat{c}+\hat{c}^\dagger)
\end{eqnarray}
where $\hat{a},\hat{b},\hat{c}$ are the (displaced) annihilation operators for the optical, microwave and mechanical  resonators respectively, the optical and microwave driving field amplitudes are $E_c,E_w$ respectively. The microwave and the optical field must be phase-locked, which can be realized by means of frequency-comb techniques \cite{Holzwarth2000}; varying this relative phase is equivalent to a local unitary operation which does not modify the entanglement between the two fields, and therefore we have chosen such a phase equal to zero, and taken $\alpha_s$ and $\beta_s$ real and positive.

Before we present the full analysis, including damping, we can illustrate the key principle. As we describe below, the mechanical resonator mediates an effective retarded interaction between the optical and cavity modes which is responsible for: i) cavity frequency shift and single mode squeezing for both modes; ii) excitation transfer between the two modes; iii) two-mode squeezing between optical and microwave photons. One can resonantly select one of these processes by appropriately adjusting the cavity detunings. For example, the state transfer schemes of Refs.~\cite{Tian2010,Taylor2011,Tian2012,Wang2012} chooses equal detunings $\Delta_c = \Delta_w $. Instead here we choose opposite detunings $\Delta_c = -\Delta_w \equiv \Delta \simeq \omega_m$, and assume the regime of fast mechanical oscillations, $\Delta \sim \omega_m \gg G_c,G_w,\kappa_c,\kappa_w$, so that we are in the resolved sideband regime for both cavities, with red sideband driving for the optical cavity and blue sideband driving for the microwave cavity. This choice allows us to neglect the fast oscillating terms at $\sim \pm 2 \Delta $.     In the case $\Delta_c = -\Delta_w \equiv \Delta \simeq \omega_m$ we can approximate the Hamiltonian by
\begin{equation}
\label{approx-ham}
H_a  =  -\hbar G_{c}(\hat{a}^\dagger\hat{c}+\hat{a}\hat{c}^\dagger) -\hbar G_{w}(\hat{b}\hat{c}+\hat{b}^{\dagger}\hat{c}^\dagger)
\end{equation}
The second term alone is responsible for entangling the microwave resonator with the mechanical resonator, while the first term alone  exchanges the states of
optical and mechanical resonator. If these terms are acting together, we anticipate a regime in which the optical and microwave resonators become entangled thereby  enabling
a continuous variable teleportation protocol to be implemented. This is confirmed in a more detailed analysis including damping.

We include damping and thermal noise by  adopting a quantum Langevin equation (QLE) in which we add to the Heisenberg equations mechanical damping with rate $\gamma_m$, the
quantum Brownian noise acting on the MR $\hat{\xi}(t)$, cavity decay rates $\kappa_c, \kappa_w$, and the optical and microwave input noises $\hat{a}_{in}(t)$ and $\hat{b}_{in}(t)$ \cite{Barzanjeh2011a}.

We now see when the proposed device behaves as a parametric oscillator involving an optical and a microwave mode. It is convenient to move to the interaction picture with respect to $\hat{H}_{\Delta}=\hbar \Delta_{c} \hat{a}^{\dagger} \hat{a}+\hbar \Delta_{w} \hat{b}^{\dagger} \hat{b}$, formally solve the dynamics of the mechanical resonator and insert this formal solution into the dynamical equations of the two modes. This gives a direct dynamical interaction between the optical and microwave resonators in terms of a convolution integral (here denoted with $*$) with the mechanical susceptibility $ \chi_M(t) = e^{-\gamma_m t/2} \sin \omega_m t$~\cite{Barzanjeh2011a},
\begin{subequations}\label{qlesappr}
\begin{eqnarray}
&&\dot{\hat{a}} =-\kappa_c \hat{a}(t)+e^{i\Delta_c t}\left\{\sqrt{2\kappa_c}\hat{a}_{in}(t)\right.\\
& &  \left.+\frac{i}{2}\left[\chi_M *\left(G_c \hat{\xi} +G_c^2 \hat{X}_a+G_c G_w\hat{X}_b\right)\right](t)\right\},\nonumber \\
&& \dot{\hat{b}}=-\kappa_w  \hat{b}(t)+e^{i\Delta_w t}\left\{\sqrt{2\kappa_w}\hat{b}_{in}(t)\right.\\
 &&\left.+\frac{i}{2}\left[\chi_M * \left(G_w \hat{\xi} +G_w^2 \hat{X}_b+G_c G_w\hat{X}_a \right)\right](t)\right\},\nonumber
\end{eqnarray}
\end{subequations}
where the optical and microwave quadrature phase operators are defined by
$\hat{X}_a(t)  =  a(t)e^{-i\Delta_c t}+a^\dagger(t) e^{i\Delta_c t}$ and
$\hat{X}_b(t)  =  b(t)e^{-i\Delta_w t}+b^\dagger(t) e^{i\Delta_w t}$,
and where $\hat{\xi}(t)$ is a Langevin thermal force term acting on the mechanical resonator.  The mechanical system is acting like a nonlinear medium mixing the two electromagnetic fields. This is analogous to the mechanically mediated electromagnetically induced transparency for optical fields~\cite{Weis2010}.  Note that we have {\em not} made the rotating wave approximation leading to the dropping of the non resonant terms as in the approximation in Eq.~(\ref{approx-ham})

As above, we choose opposite detunings $\Delta_c = -\Delta_w \equiv \Delta \simeq \omega_m$, and assume that we are in the resolved sideband regime for both cavities. Under these conditions, the two modes undergo a retarded parametric interaction with a time-dependent coupling kernel $G_c G_w\chi_M e^{i\Delta t}/2$, which is resonantly large only if $\Delta \sim \omega_m$, because otherwise the kernel rapidly oscillates and the interaction tends to average to zero. Therefore, it is convenient to choose one of the two resonant conditions, $\Delta_w=-\Delta_c=\pm \omega_m$, that is, one cavity mode is resonant with the red sideband and the other one with the blue sideband of the respective driving field.

This argument explains how one can entangle the \emph{intracavity} microwave and optical modes \cite{Barzanjeh2011a}. However in quantum communication protocols one manipulates and entangles \emph{traveling output} electromagnetic modes. Therefore we focus on the steady state of the system formed by the two (eventually filtered) cavity outputs, one at optical and the other at microwave frequencies. When such a state is entangled, the proposed device represents an extremely robust resource for any quantum information protocol, owing to the virtually infinite entanglement lifetime.

In experiment the cavity output modes are mixed with a strong local oscillator prior to detection on a photodetector resulting in a homodyne current. This current can then be integrated over some appropriate time window.   By appropriately choosing the temporal mode functions of the local oscillator we can thus define the measurement in terms of filtered output modes.  By properly choosing the central frequency and the bandwidth of the local oscillator, one can \emph{optimally filter} the entanglement between the two output modes \cite{Genes2008b}. Other filtering methods, e.g. optoelectronic phase modulation can also be used~\cite{Milburn2008}. This is analogous to what happens in single-mode optical squeezing \cite{Walls1995}: intracavity squeezing is always limited, while one can achieve arbitrary squeezing in an appropriate narrow bandwidth of the output spectrum.

The measured cavity output modes are defined by the following bosonic annihilation operators \cite{Genes2008b}
\begin{subequations}\label{kernel}
\begin{eqnarray}
\hat{a}_c^{out}(t)= \int_{-\infty}^{t}ds g_c(t-s)\hat{a}^{out}(s),\\
\hat{b}_w^{out}(t)= \int_{-\infty}^{t}ds g_w(t-s)\hat{b}^{out}(s),
\end{eqnarray}
\end{subequations}
where $\hat{a}^{out}(t)=\sqrt{2\kappa_c}\delta \hat{a}(t)-\hat{a}^{in}(t)$, and $\hat{b}^{out}(t)=\sqrt{2\kappa_w}\delta \hat{b}(t)-\hat{b}^{in}(t)$ are the standard input-output relationships for the optical and microwave fields \cite{Walls1995}, and $g_c(t)$ and $g_w(t)$ are causal filter functions defining the output modes. In fact, $\hat{a}_c^{out}$ and $\hat{b}_w^{out}$ are standard photon annihilation operators, implying the normalization conditions $\int dt |g_c(t)|^2 =\int dt |g_w(t)|^2=1$. A simple choice is taking
$g_j(t)=\sqrt{2/\tau}\theta(t)e^{-(1/\tau+i\Omega_j) t}$, $j=c,w$,
where $\theta(t)$ is the Heaviside step function, $1/\tau$ is the bandwidth of the output modes (equal for the two modes), and $\Omega_j$ is the central frequency (measured with respect to the frequency of the corresponding driving field).

The stationary state of the system is a zero-mean Gaussian state because the system is driven by Markovian Gaussian noises $\xi(t)$, $a^{in}$ and $b^{in}$, and we are considering the linearized dynamics of the quantum fluctuations around the semiclassical fixed point \cite{Barzanjeh2011a}. Therefore it is straightforward to quantify its entanglement by computing the corresponding logarithmic negativity $E_N$ \cite{Eisert2001} (see Supplementary Material).
As expected, we find large entanglement (much larger than that between \emph{intracavity} modes \cite{Barzanjeh2011a}) in the limit of narrow-band output modes of the microwave and optical cavities, under the resonant condition $\Delta_w=-\Delta_c=\omega_m$ (see Fig.~2). Large entanglement is achieved only around $\Omega_w=\omega_m$, for fixed central frequency of the optical output mode $\Omega_c=-\omega_m$, and for increasingly narrow output bandwidths. This means that the common interaction with the MR establishes quantum correlations between the microwave and optical cavity outputs, which are strongest between the Fourier components \emph{exactly} at resonance with the respective cavity field.

\begin{figure}[ht]
   \centering
   \includegraphics[width=.45\textwidth]{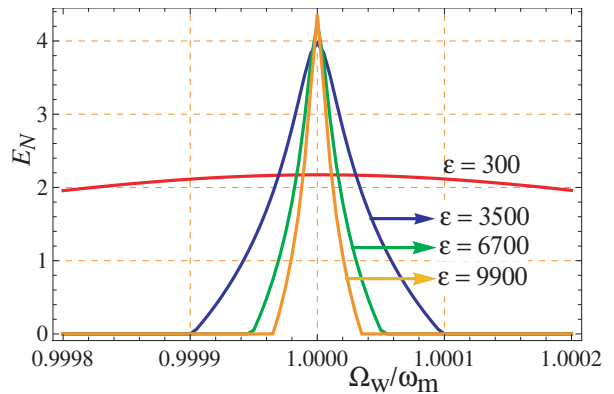} 
   \caption{\small{$E_{N}$ at four different values of the normalized inverse of the bandwidth $\epsilon=\tau\omega_m$ vs the normalized frequency $\Omega_w/\omega_{m}$, at fixed central frequency of the optical output mode $\Omega_c=-\omega_m$. The optical and microwave cavities detunings have been fixed at $\Delta_c = -\Delta_w =-\omega_m$, while the other parameters are $\omega_m/2\pi=10$ MHz, $Q\equiv \omega_m/\gamma_m=1.5\times 10^5$, $\omega_w/2\pi=10$ GHz, $\kappa_w=0.04\omega_m$, $P_w=42$ mW, $m=10$ ng, $T=15$ mK, $d=100$ nm, $\mu=0.013$, where $d$ and $\mu$ are parameters of the equivalent capacitor defined in the Supplementary materials. This set of parameters is analogous to that of Teufel et al.~\protect\cite{Teufel2011} for the MC and MR, except that we have considered a lower mechanical quality factor, and a heavier mass, in order to take into account the presence of the coating. We have then assumed an OC of length $L=1$ mm and damping rate $\kappa_c=0.04\omega_m$, driven by a laser with wavelength $\lambda_{0c}=810$ nm and power $P_c=3.4$ mW.}
   }
   \label{fig:2}
\end{figure}

Equivalently, entanglement is maximum when narrow-band blue-detuned microwave and red-detuned optical output fields are selected, i.e., $\Omega_w=\Delta_w=-\Delta_c=-\Omega_c=\omega_m$. Fig.~\ref{fig:2} refers to a parameter set representing a feasible extension of the scheme of Ref.~\cite{Teufel2011}, i.e., a lumped-element superconducting circuit with a free standing drum-head capacitor which is then optically coated to form a micromirror of an additional optical Fabry-Perot cavity. Fig.~\ref{fig:2} shows in practice a very efficient source of two-mode squeezing, in which the idler (signal) is at \emph{optical} frequencies, and the signal (idler) is at \emph{microwave} frequencies.

Such a large stationary entanglement can be exploited for the implementation of continuous variable (CV) quantum teleportation~\cite{Braunstein1998}.
A key role is played by  two phase locked local oscillator fields, one for the optical output and one for the microwave output. These local oscillator fields need to be chosen in an appropriate temporal
mode to effect the required filtering to access the large steady state entanglement produced by the optomechanical interface.
An unknown state, the client (Victor) state, of the optical field is prepared and sent to Alice,
where it is mixed at a balanced beam splitter with the optical output of the device proposed here and the optical local oscillator with the appropriate temporal mode shape.

Alice performs a balanced homodyne detection at each output port of the beam splitter, effecting a joint measurement of two temporally filtered quadrature phase operators and sends the results of her measurements to Bob as
a classical current.   Bob uses the measured homodyne current sent from Alice to effect an appropriate conditional, coherent displacement of the microwave field at his location. This is done by mixing the microwave local oscillator and the microwave output from the interface on an almost perfectly reflecting beam spitter, with the phase and amplitude of the local oscillator chosen according to the measurement current received from Alice. By exploiting only homodyne measurements and coherent field displacements conditioned on classical communication the strong entanglement realised by the proposed optomechanical hybrid device a quantum state of an optical field can be teleported onto the quantum sate of a microwave field. The entire protocol can be reversed, i.e. joint measurement can be performed at the microwave end and conditional coherent displacements performed at the optical end.

The quality of the proposed teleportation protocol is quantified by the fidelity $F$ which, in the case of a pure input state $|\psi_{in}\rangle$ at Victor site, is given by $F=\langle\psi_{in}|\rho_{out}|\psi_{in}\rangle$, where $\rho_{out}$ is the output state at Bob site after the conditional displacement. In terms of the Wigner characteristic functions of the input state $\Phi_{in}(\alpha)$ and of the entangled channel $\Phi_{ch}(\alpha,\beta)$, one has $F=\pi^{-1}\int d^2\alpha |\Phi_{in}(\alpha)|^2 \Phi_{ch}(\alpha^*,\alpha)^*$~\cite{Pirandola2006a}. We consider a highly non-classical input state at Victor site, an even Schr\"odinger cat state $|\psi\rangle=N(|\alpha\rangle+|-\alpha\rangle)$, where $N=\left\{2+2\exp\left[-2\alpha^2\right]\right\}^{-1/2}$. Various possible threshold fidelities have been suggested for unambiguously distinguishing a successful quantum teleportation from the best classical state transfer strategy \cite{Lee2011}. In the case of non-classical states however one can adopt the so-called ``no-cloning threshold'' $F_{th}=2/3$ \cite{Grosshans2001}, which has the following property: any non-classical input state (i.e., possessing a negative Wigner function) remains non-classical at the end of the teleportation protocol, if and only if $F>F_{th}$ \cite{Ban2004}.

\begin{figure}[ht]
   \centering
  \includegraphics[width=.45\textwidth]{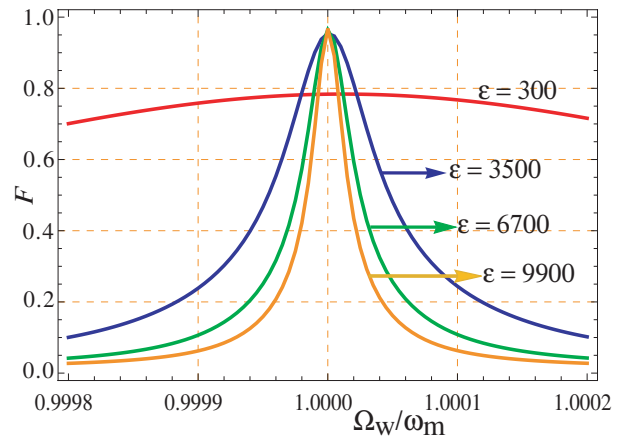} 
   \caption{Teleportation fidelity $F$ at four different values of $\epsilon=\tau\omega_m$ vs $\Omega_w/\omega_m$ and for Schr\"odinger cat-state amplitude $\alpha=1$. The other parameters are as in Fig.~\protect\ref{fig:2}.
   }
   \label{fig:3}
\end{figure}

\begin{figure}[ht]
   \centering
   \includegraphics[width=.45\textwidth]{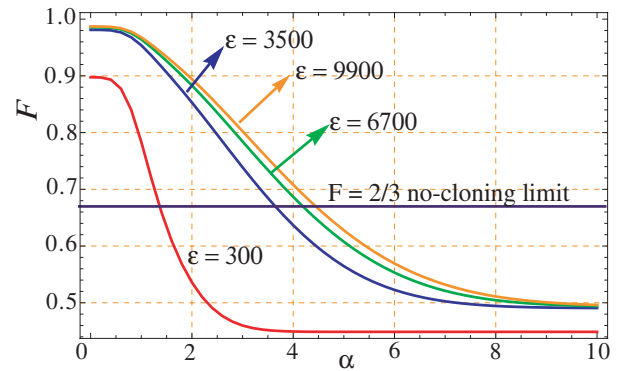} 
   \caption{Teleportation fidelity $F$ at four different values of $\epsilon=\tau\omega_m$ vs the Schr\"{o}dinger cat-states amplitude $\alpha$, at a fixed central frequency of the microwave output mode $\Omega_w= \omega_m$ and fixed temperature $T=15$ mK. The other parameters are as in Fig.~\protect\ref{fig:2} .
   }
   \label{fig:4}
\end{figure}

$F$ for the even Schr\"odinger cat state can be evaluated explicitly (see Supplementary Material), and the results are shown in Figs.~\ref{fig:3} and \ref{fig:4}. $F$ shows the same behavior of $E_N$: this fact, although intuitive, is not generally true because $F$ depends upon the protocol details, and is not invariant under local unitary transformations, i.e., it is not an entanglement measure. The similar behavior of $E_N$ and $F$ is here a consequence of the fact that Alice's choice of measuring $\hat X_+$ and $\hat Y_-$ for her Bell measurement is close to be optimal for the Gaussian entangled state shared by the two parties, because it exploits the sub-shot noise variance of $\hat{X}_{c}^{out}+ \hat{X}_{w}^{out}$ and $\hat{Y}_{c}^{out}- \hat{Y}_{w}^{out}$. In fact, the selected narrow-band microwave and optical output modes possess Einstein-Podolsky-Rosen (EPR) correlations that can be immediately exploited for teleportation without any need for local optimizations such as those discussed in Refs.~\cite{Fiurasek2002,Mari2008}. This is confirmed by the fact that $F$ is very close to the optimal upper bound achievable for a given $E_N$, $F_{opt}=\left(1+e^{-E_N}\right)^{-1}$ \cite{Mari2008,Adesso2005}. Finally Fig.~\ref{fig:4} shows $F$ versus the amplitude of the cat state $\alpha$ for different values of the inverse bandwidth $\tau$ when the optimal resonance condition $\Omega_w=\omega_m$ is taken.

The teleportation protocol can be reversed, and the role of the optical and microwave output fields can be exchanged, by exploiting the symmetry of the effective parametric interaction mediated by the MR. This means that by exchanging in Fig.~1 the roles of Alice and Bob, one can teleport the state of an input microwave field onto the output optical field at Alice site. This means mixing and homodyning microwave fields for the Bell measurements at Bob's location, and conditionally displacing the state of the optical field at Alice's location.
The only problem in this reversal is technical, because it is presently difficult to achieve as high an efficiency for homodyne detection of microwave fields as it is for optical fields. However single-photon counter detectors at microwave frequencies are under development, and therefore there is no serious limitation for implementing CV Bell measurements at microwave wavelengths.


\emph{Acknowledgements} This work has been supported by the European Commission through the FP-7 FET-Open projects MINOS and HIP. GJM was supported by the Australian Research Council grant CE110001013.

\bibliography{optomechanics-telep}

\begin{appendix}

\section{Supplementary information for ``Reversible optical to microwave quantum interface''}

\subsection{Derivation of the nonlinear Quantum langevin Equations and their linearization}

The Hamiltonian of the system under study is given by
\begin{eqnarray}\label{ham0SM}
H&=&\frac{p_x^2}{2m}+\frac{m\omega^2_m
x^2}{2}+\frac{\Phi^2}{2L}+\frac{Q^2}{2[C+C_0(x)]}-e(t)Q\\
&&+\hbar\omega_c a^{\dagger}a-\hbar G_{0c}a^{\dagger}ax+i\hbar
E_c(a^{\dagger}e^{-i\omega_{0c}t}-a e^{i\omega_{0c}t}).\nonumber
\end{eqnarray}
where $(x,p_x)$ are the canonical position and momentum of a
MR with frequency $\omega_m$, $(\Phi,Q)$ are the
canonical coordinates for the MC describing the flux
through an equivalent inductor $L$ and the charge on an equivalent
capacitor $C$, respectively, $(a,a^{\dagger})$ show the annihilation
and creation operators of the OC
mode($[a,a^{\dagger}]=1$), $E_c=\sqrt{2P_c\kappa_c/\hbar\omega_{0c}}$
is related to input driving laser, where $P_c$ is the power of the
input laser and $\kappa_c$ describes the damping rate of the optical
cavity. $G_{0c}=(\omega_c/{\cal L})\sqrt{\hbar/m\omega_m}$ gives the optomechanical coupling rate, with $m$ the
effective mass of mechanical mode, and ${\cal L}$ the length
of the optical Fabry-Perot cavity, while the coherent
driving of the MC with damping rate $\kappa_w$ is given
by electric potential
$e(t)=-i\sqrt{2\hbar\omega_wL}E_w(e^{i\omega_{0w}t}-e^{-i\omega_{0w}t})$.
The MR is coupled to the microwave cavity because the capacity of the latter is a function of the resonator displacement,
$C_0(x)$. We expand this function around the equilibrium position of the resonator corresponding to a separation
$d$ between the plates of the capacitor, with corresponding bare capacitance $C_0$,
$C_0(x)=C_0[1+x(t)/d]$. Expanding the capacitive energy as a
Taylor series, we find to first order,
\begin{eqnarray}\label{2}
 \frac{Q^2}{2[C+C_0(x)]}=\frac{Q^2}{2C_{\Sigma}}-\frac{\mu}{2d
C_{\Sigma}}x(t)Q^2,
\end{eqnarray}
where $C_\Sigma=C+C_0$ and $\mu=C_0/C_\Sigma$.
The Hamiltonian of Eq.~(\eqn{ham0SM}) can be rewritten in the terms of the
raising and lowering operators of the MC field $b,
b^{\dagger}$($[b,b^{\dagger}]=1$) and the dimensionless position and
momentum operators of the MR, $\hat q$, $\hat p$ ($[\hat q,\hat
p]=i$), as
\begin{eqnarray}\label{ham1}
H&=&\hbar \omega_w b^{\dagger}b+\hbar \omega_c
a^{\dagger}a+\frac{\hbar \omega_m}{2}(\hat p^2+\hat q^2)-\frac{\hbar
G_{0w}}{2}\hat q(b+b^{\dagger})^2\nonumber\\
&&-\hbar G_{0c}\hat q a^{\dagger}a-i\hbar
E_w(e^{i\omega_{0w}t}-e^{-i\omega_{0w}t})(b+b^{\dagger})\nonumber\\
&&+i\hbar E_c(a^{\dagger}e^{-i\omega_{0c}t}-a e^{i\omega_{0c}t}),
\end{eqnarray}
where
\begin{eqnarray}\label{4}
b &=& \sqrt{\frac{\omega_w L}{2\hbar}}\hat Q+\frac{i}{\sqrt{2\hbar
\omega_w L}}\hat \Phi,\\
\hat q&=&\sqrt{\frac{m\omega_m}{\hbar}}\hat x,\;\;\;\;
\hat p=\frac{\hat p_{x}}{\sqrt{\hbar m\omega_m}},\\
G_{0w}&=&\frac{\mu \omega_w}{2d}\sqrt{\frac{\hbar}{m\omega_m}}.
\end{eqnarray}
It is then convenient to adopt the interaction picture with respect to
$H_0=\hbar \omega_{0w}b^{\dagger}b+\hbar \omega_{0c}a^{\dagger}a$, and neglect fast oscillating terms at $\pm2\omega_{0w}, \pm 2\omega_{0c}$, so that the system Hamiltonian becomes
\begin{eqnarray}\label{ham2}
H&=&\hbar \Delta_{0w} b^{\dagger}b+\hbar \Delta_{0c}
a^{\dagger}a+\frac{\hbar \omega_m}{2}(\hat p^2+\hat q^2)-\hbar
G_{0w}\hat qb^{\dagger}b\nonumber\\
&&-\hbar G_{0c}\hat q a^{\dagger}a-i\hbar E_w(b-b^{\dagger})+i\hbar
E_c(a^{\dagger}-a).
\end{eqnarray}
However the dynamics of the three modes is also affected by damping and noise processes, due to the fact that each of them interacts with its own environment. We can describe them adopting a Quantum Langevin Equation (QLE) treatment in which the Heisenberg equations for the system operators associated with Eq.~(\ref{ham2}) are supplemented with damping and noise terms. The resulting nonlinear QLEs are given by
\begin{eqnarray}\label{lan1}
\dot{q}&=&\omega_m p,\\
\dot{p}&=&-\omega_m q-\gamma_m
p+G_{0c}a^{\dagger}a+G_{0w}b^{\dagger}b+\xi, \label{lan2}\\
\dot{a}&=&-(\kappa_c+i\Delta_{0c})a+iG_{0c}q
a+E_c+\sqrt{2\kappa_c}a_{in}, \label{lan3}\\
\dot{b}&=&-(\kappa_w+i\Delta_{0w})b+iG_{0w}q
b+E_w+\sqrt{2\kappa_w}b_{in}, \label{lan4}
\end{eqnarray}
where $\gamma_m$ is the mechanical damping rate and
$\xi(t)$ is the quantum Brownian noise acting on the MR, with
correlation function~\cite{Giovannetti2001}
\begin{equation}\label{nois1}
\langle\xi(t)\xi(t')\rangle=\frac{\gamma_m}{\omega_m}\int
\frac{d\omega}{2\pi}
e^{-i\omega(t-t')}\omega\left[\coth\left(\frac{\hbar \omega}{2k_B
T}\right)+1\right],
\end{equation}
where $k_B$ is the Boltzmann constant and $T$ is the temperature
of the reservoir of the mechanical resonator. We have also introduced the
optical and microwave input noises, respectively given by $a_{in}(t)$ and $b_{in}(t)$, obeying the following correlation functions \cite{Gardiner2000}
\begin{eqnarray}\label{coropt}
\langle
a_{in}(t)a^{\dagger}_{in}(t')\rangle&=&[N(\omega_c)+1]\delta(t-t'),\\
\langle a^{\dagger}_{in}(t)a_{in}(t')\rangle&=&N(\omega_c)\delta(t-t'),\\
\label{cormicro}
\langle b_{in}(t)b^{\dagger}_{in}(t')\rangle&=&[N(\omega_w)+1]\delta(t-t'),\\
\langle b^{\dagger}_{in}(t)b_{in}(t')\rangle&=&N(\omega_w)\delta(t-t'),
\end{eqnarray}
where $N(\omega_c)=[\mathrm{exp}(\hbar \omega_c/k_B T)-1]^{-1}$ and
$N(\omega_w)=[\mathrm{exp}(\hbar \omega_w/k_B T)-1]^{-1}$ are the equilibrium
mean thermal photon number of the optical and microwave field,
respectively. One can safely assume $N(\omega_c)\approx0$ since $\hbar \omega_c/k_B T \gg 1$ at optical frequencies, while thermal microwave photons cannot be neglected in general, even at very low temperatures.

The physically relevant regime is when both modes are strongly driven and one has intense intracavity optical and microwave fields. In this case the interested quantum dynamics is retained by the quantum fluctuations around the classical steady state of the system, which must be a stable fixed point of the classical dynamics. For this purpose, one can write $a=\alpha_s+\delta a$, $b=\beta_s+\delta b$, $p=p_s+\delta p$, and $q=q_s+\delta q$ and by solving the classical equations for the steady state values, one finds that the fixed point is given by $p_s=0,\,q_s=\left[G_{0c}|\alpha_s|^2+G_{0w}|\beta_s|^2\right]/\omega_m$, where $\alpha_s$ and $\beta_s$ are the solutions of the nonlinear equations $|E_c|^2=|\alpha_s|^2[\kappa_c^2+\Delta_c^2]$ and $|E_w|^2=|\beta_s|^2[\kappa_w^2+\Delta_w^2]$, where $\Delta_i=\Delta_{0i}-G_{0i}q_s$, $i=c,w$, describe the effective detunings of the cavity fields. The stability of this fixed point can be verified for example by using the Routh-Hurwitz criterion, and we have always considered such a stable regime in this paper. The exact QLE can be safely linearized when the intracavity fields are very intense, i.e., $|\alpha_s|\gg 1$ and $|\beta_s|\gg 1$ and under these conditions, one obtains the following linear QLE for the quantum fluctuations of the tripartite system
\begin{subequations}\label{qles2SM}
\begin{eqnarray}
\delta \dot{q}&=&\omega_m \delta p,\\
\delta \dot{p}&=&-\omega_m \delta q-\gamma_m p+G_{0c}\alpha_s (\delta
a^{\dagger}+\delta a)\nonumber \\
&&+G_{0w}\beta_s (\delta b^{\dagger}+\delta b)+\xi,\\
\delta \dot{a}&=&-(\kappa_c+i\Delta_c)\delta a+i G_{0c}\alpha_s \delta
q+\sqrt{2\kappa_c}a_{in},\\
\delta \dot{b}&=&-(\kappa_w+i\Delta_w)\delta b+i G_{0w}\beta_s \delta
q+\sqrt{2\kappa_w}b_{in},
\end{eqnarray}
\end{subequations}
where we have chosen the phase references so that $\alpha_s$ and
$\beta_s$ can be taken real and positive.

\subsection{Derivation of the stationary covariance matrix and the associated entanglement}

The linearized QLE derived above can be rewritten in terms of the amplitude and phase quadrature fluctuations operators of the two electromagnetic fields, $\delta X_c=(\delta a+\delta a^{\dagger})/\sqrt{2}$, $\delta Y_c=(\delta a-\delta a^{\dagger})/i\sqrt{2}$, $\delta X_w=(\delta b+\delta b^{\dagger})/\sqrt{2}$, and $\delta Y_w=(\delta b-\delta b^{\dagger})/i\sqrt{2}$, and of the corresponding Hermitian input noise operators $X^{in}_{c}=(a_{in}+ a_{in}^{\dagger})/\sqrt{2}$, $Y^{in}_{c}=( a_{in}-a_{in}^{\dagger})/i\sqrt{2}$, $X^{in}_{w}=( b_{in}+ b_{in}^{\dagger})/\sqrt{2}$, $ Y^{in}_{w}=( b_{in}- b_{in}^{\dagger})/i\sqrt{2}$. Using these definitions,
Eqs.~(\ref{qles2SM}) can be written in matrix form as $\dot u(t)=A u(t)+n(t)$, where $u^T(t)=[\delta q(t),\delta p(t),\delta X_w(t),\delta Y_w(t),\delta X_c(t),\delta Y_c(t)]^T$, $n(t)=[0,\xi(t),\sqrt{2\kappa_w} X^{in}_w,\sqrt{2\kappa_w} Y^{in}_w,\sqrt{2\kappa_c} X^{in}_c,\sqrt{2\kappa_c} Y^{in}_c]^T$ is a vector of noise operators, and $A$ is the drift matrix
\begin{equation}\label{driftA}
A = \left( {\begin{array}{*{20}c}
   0 & {\omega _m } & 0 & 0 & 0 & 0  \\
   { - \omega _m } & { - \gamma _m } & {G_w } & 0 & {G_c } & 0  \\
   0 & 0 & { - \kappa _w } & {\Delta _w } & 0 & 0  \\
   {G_w } & 0 & { - \Delta_w } & { - \kappa _w } & 0 & 0  \\
   0 & 0 & 0 & 0 & { - \kappa _c } & {\Delta _c }  \\
   {G_c } & 0 & 0 & 0 & { - \Delta _c } & { - \kappa _c }  \\
\end{array}} \right).
\end{equation}
We are interested however in the \emph{output} quadrature fluctuations
\begin{equation}\label{vectorSM}
u^{out}_i(t)=[\delta q(t),\delta p(t), X^{out}_w(t),Y^{out}_w(t), X^{out}_c(t),Y^{out}_c(t)]^T,
\end{equation}
and in the associated covariance matrix in the stationary state of the system,
\begin{equation}\label{cor1SM}
V^{out}_{ij}(\infty)=\lim_{t \to \infty}\frac{1}{2}<u^{out}_i(t)u^{out}_j(t)+u^{out}_j(t)u^{out}_i(t)>,
\end{equation}
where we have defined the annihilation operator of the output modes
\begin{eqnarray}\label{kernel1}
a_c^{out}(t)&=& \int_{\infty}^{t}ds g_c(t-s)a^{out}(s),\\
b_w^{out}(t)&=& \int_{\infty}^{t}ds g_w(t-s)b^{out}(s),\label{kernel2}
\end{eqnarray}
where $a^{out}(t)=\sqrt{2\kappa_c}\delta a(t)-a^{in}(t)$, and $b^{out}(t)=\sqrt{2\kappa_w}\delta b(t)-b^{in}(t)$ are the standard input-output relationships for the optical and microwave fields \cite{Gardiner2000}, and $g_c(t)$ and $g_w(t)$ are causal filter functions defining the output modes.
Following the derivation of Ref.~\cite{Genes2008b}, and passing to the frequency domain, one arrives at the following general expression for
the stationary output correlation matrix~\cite{Genes2008b}
\begin{align}\label{vmat}
V^{out}(\infty)=\int &d\omega\tilde{T}(\omega)\Big(\tilde M^{ext}(\omega)+P_{out}\Big)\nonumber\\
&\times D_{ext}\Big(\tilde M^{ext}(\omega)^{\dagger}+P_{out}\Big)\tilde{T}^{\dagger}(\omega),
\end{align}
where $\tilde{T}(\omega)$ is the Fourier transforms of
\begin{align}
T(t) = \left( {\begin{array}{*{20}c}
   {\delta (t)} & 0 & 0 & 0 & 0 & 0  \\
   0 & {\delta (t)} & 0 & 0 & 0 & 0  \\
   0 & 0 & R_w & -I_w & 0 & 0  \\
   0 & 0 & I_w & R_w & 0 & 0  \\
   0 & 0 & 0 & 0 & R_c & -I_c  \\
   0 & 0 & 0 & 0 & I_c & R_c  \\
\end{array}} \right),
\end{align}
$\tilde M^{ext}(\omega)=(i\omega+A)^{-1}$, $P_{out}=\mathrm{Diag}[0,0,1/2k_w,1/2k_w,1/2k_c,1/2k_c]$, $D^{ext}=\mathrm{Diag}[0,\gamma_m(2\bar{n}_b+1),2\kappa_w(2N(\omega_w)+1),2\kappa_w(2N(\omega_w)+1),2\kappa_c,2\kappa_c]
$, $R_j=\sqrt{2\kappa_j}{\rm Re}[g_j(t)]$, $I_j=\sqrt{2\kappa_j}{\rm Im}[g_j(t)]\,\,(j=c,w)$.

In order to establish the conditions under which the output of optical and microwave modes are entangled, we
consider the logarithmic negativity $E_N$, which can be defined as~\cite{Eisert2001}
\begin{equation}\label{loga}
E_N={\rm Max}[0,-\ln2\eta],
\end{equation}
where $\eta \equiv2^{-1/2}\left[\Sigma(V')-\sqrt{\Sigma(V')^2-4 \mathrm{det} V'}\right]^{1/2}$, with $V'$ the reduced CM of the two output modes, which is written in terms of $2\times2$ block matrix form as
\begin{equation}\label{loga2}
V'=\left(
     \begin{array}{cc}
       B & C \\
       C^T & B' \\
     \end{array}
   \right),
\end{equation}
and with $\Sigma(V')\equiv \mathrm{det} B+\mathrm{det} B'-2\mathrm{det} C$.

\subsection{Derivation of the teleportation fidelity for an input Schr\"odinger cat state}

We consider as input state provided by Victor the following even Schr\"odinger cat state $|\psi\rangle=N(|\alpha\rangle+|-\alpha\rangle)$, where $N=\left\{2+2\exp\left[-2\alpha^2\right]\right\}^{-1/2}$ and $\alpha$ is taken real (for $\alpha \to 0$ the state reduces the vacuum state). From the definition and using the Wigner characteristic function \cite{Gardiner2000}, one can express in general the fidelity in the following form \cite{Pirandola2006}
\begin{eqnarray}\label{fidchar}
F=\pi^{-1}\int d^2\eta|\phi^{in}(\eta)|^2[\Phi^{ch}(\eta^*,\eta)]^*,
\end{eqnarray}
where
\begin{eqnarray}
&&\phi^{in}(\eta)=N^2 e^{-\frac{|\eta|^2}{2}}\left\{\exp\left[\alpha(\eta-\eta^*)\right]+\exp\left[-\alpha(\eta-\eta^*)\right]\right.\nonumber \\
&&\left.+2\exp\left[-2\alpha^2\right]\cosh\left[\alpha(\eta+\eta^*)\right]\right\},
\end{eqnarray}
is the Wigner characteristic function of the input cat state, and $\Phi^{ch}(\eta_a,\eta_b)$ is the Wigner characteristic function of the generic bipartite entangled state shared by Alice and Bob.

We now restrict to the case under study, in which the entangled resource shared by Alice and Bob is a CV Gaussian state with zero mean, associated with the reduced state of the microwave and optical output modes. In such a case one has $\Phi^{ch}(\eta_a,\eta_b)=\exp\left\{-\vec\xi^T V'\vec\xi\right\}$, where $V'$ is the $4\times 4$ matrix given by Eq.~(\ref{loga2}) and derived from the $6\times 6$ matrix of Eq.~(\ref{vmat}), and $\vec\xi^T=(\eta_a^I,-\eta_a^R,\eta_b^I,-\eta_b^R)$ is a four dimensional vector extracted from the complex variables $\eta_a = \eta_a^R+i\eta_a^I$ and $\eta_a = \eta_b^R+i\eta_b^I$ \cite{Pirandola2006}.

Using these expression the integral of Eq.~(\ref{fidchar}) becomes an involved Gaussian integral, which can be explicitly calculated. One finally gets
\begin{eqnarray}
&&F=\frac{N^4}{\sqrt{\mathrm{det}\Gamma}}\left\{e^{-Q_1(\alpha)}+e^{-Q_1(-\alpha)}\right.\\
&&\left.+e^{-4\alpha^2}\left[e^{-Q_2(\alpha)}
+e^{-Q_2(-\alpha)}\right]\right. \nonumber \\
&&\left.+2e^{-2\alpha^2} \left[e^{-Q_3(\alpha)}+e^{-Q_3(-\alpha)}+e^{-Q_4(\alpha)}+e^{-Q_4(-\alpha)}\right]\right.\nonumber \\
&&\left.+ 2e^{-4\alpha^2}+2\right\},
\end{eqnarray}
where $\Gamma$ is the $2 \times 2$ matrix given by
\begin{equation}
\Gamma= I+ZBZ+ZC+C^TZ+B',
\end{equation}
where $I$ is the identity matrix and $Z={\rm Diag}(1,-1)$. Moreover $Q_i(\alpha)={\vec h}_i^T(\alpha)\Gamma^{-1}{\vec h}_i(\alpha)$, with
\begin{align}
{\vec h}_1(\alpha)=\left(
                     \begin{array}{c}
                       2\alpha \\
                       0 \\
                     \end{array}
                   \right),\,\,{\vec h}_2(\alpha)=\left(
                     \begin{array}{c}
                       0 \\
                       2i\alpha \\
                     \end{array}
                   \right),\nonumber\\
{\vec h}_3(\alpha)=\left(
                     \begin{array}{c}
                       \alpha \\
                       i\alpha \\
                     \end{array}
                   \right),\,\,{\vec h}_4(\alpha)=\left(
                     \begin{array}{c}
                       \alpha \\
                       -i\alpha \\
                     \end{array}
                   \right).
\end{align}
In the limit $\alpha\to 0$ the input state becomes the vacuum state, $Q_j(\alpha)\to 0$, and $F$ tends to the usual expression for the teleportation of a coherent state, $F =  [\mathrm{det}\Gamma]^{-1/2}$ \cite{Fiurasek2002,Pirandola2006}.

\subsection{Reversibility of the quantum interface}

The proposed quantum interface may operate also in the reversed direction, i.e., it can be used to teleport an unknown state of an input \emph{microwave} field received at Bob site onto an output optical field at Alice site. Assuming that the homodyne detections on the microwave fields mixed at the beam splitter can be performed with the same high quantum efficiency that can be achieved with the optical fields at Alice site, verifying the reversible property of the proposed device is equivalent to verify that one can achieve high teleportation fidelity when the roles of the output microwave and optical fields are exchanged. In mathematical terms this is equivalent to apply the same formula above to a new $2 \times 2$ CM, $V"$ obtained from Eq.~(\ref{loga2}) by exchanging $B \leftrightarrow B'$ and $C \leftrightarrow C^{T}$. The corresponding result is shown in Fig.~1, where the teleportation fidelity of a cat state with $\alpha=1$ of an input microwave field under the same set of parameters of Fig.~3 of the paper is shown. $F$ vs the normalized central frequency of the optical output field $\Omega_c/\omega_m$ is plotted at four different values of $\epsilon=\tau\omega_m$. The results are analogous to Fig.~3, clearly showing that, as long as high-efficiency homodyne detection of microwave fields is possible, the proposed quantum interface is able to operate in both directions converting quantum states of optical field to microwave frequencies and viceversa. The highest fidelity is achieved under the same conditions in both cases, i.e., when $\Omega_w=\Delta_w=-\Delta_c=-\Omega_c=\omega_m$.

\begin{figure}[ht]
   \centering
  \includegraphics[width=.45\textwidth]{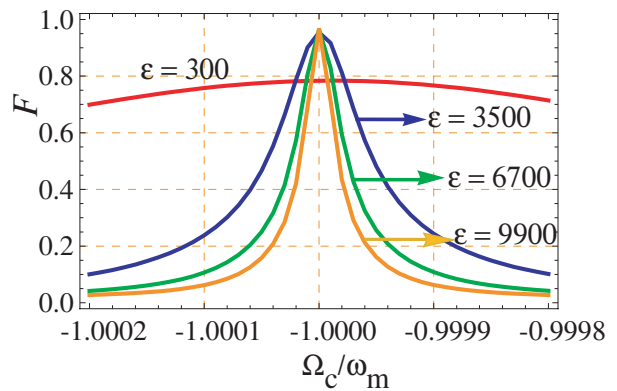} 
   \caption{Teleportation fidelity $F$ of an unknown Schr\"odinger cat-state of an input microwave field with amplitude $\alpha=1$ at four different values of $\epsilon=\tau\omega_m$ vs $\Omega_c/\omega_m$ and at fixed central frequency of the microwave mode $\Omega_w=\omega_m$. The other parameters are as in the paper: the optical and microwave cavities detunings have been fixed at $\Delta_c = -\Delta_w =-\omega_m$, while the other parameters are $\omega_m/2\pi=10$ MHz, $Q\equiv \omega_m/\gamma_m=1.5\times 10^5$, $\omega_w/2\pi=10$ GHz, $\kappa_w=0.04\omega_m$, $P_w=42$ mW, $m=10$ ng, $T=15$ mK, $d=100$ nm, $\mu=0.013$. We have also taken an OC of length $L=1$ mm and damping rate $\kappa_c=0.04\omega_m$, driven by a laser with wavelength $\lambda_{0c}=810$ nm and power $P_c=3.4$ mW.}
   \label{revers}
\end{figure}

\end{appendix}

\end{document}